# Protocol for Discovering Convergent Molecular Networks across Multi-Omics Datasets Using Network Propagation

Charles Ochieng' Olwal,[1,2,3,5]* Declan M. Winters,[1,2,3] Immy A. Ashley,[1,2,3] Valerie Vandersall,[1,2,3] Amberin Rahman,[1,2,3] Alexandra Lindsey Djomkam Zune,[4] Kyla F. Dangerfield,[1,2,3] Yennifer Delgado,[1,2,3] Sara K. Makanani,[1,2,3] Sophie F. Blanc,[1,2,3] Shipra Sharma,[1,2,3] and Mehdi Bouhaddou[1,2,3,6]*

[1]Department of Microbiology, Immunology, and Molecular Genetics (MIMG), University of California, Los Angeles, LA, USA.
[2]Institute for Quantitative and Computational Biosciences (QCBio), University of California, Los Angeles, LA, USA
[3]Molecular Biology Institute, University of California, Los Angeles, LA, USA
[4]West African Centre for Cell Biology of Infectious Pathogens (WACCBIP), Department of Biochemistry, Cell and Molecular Biology, College of Basic and Applied Sciences, University of Ghana, Accra, Ghana
[5]Technical contact
[6]Lead contact

* Correspondence should be addressed to Charles O. Olwal (charlesolwal26@ucla.edu) and Mehdi Bouhaddou (bouhaddou@ucla.edu).

## Summary
Phenotypic outcomes arise from coordinated changes across molecular layers. Although multi-omics data capture complementary molecular signals, integrating them into coherent biological mechanisms remains challenging due to limited overlap at the gene level. Network propagation maps significant gene hits from each dataset onto a prior knowledge molecular interaction network and diffuses their signals to neighboring genes. Mathematical integration of propagated networks, followed by clustering and pathway enrichment, reveals shared molecular networks, pinpointing the core biology captured across orthogonal molecular layers.

For complete details on the use and execution of this protocol, please refer to Olwal et al.[1,2.]

## Graphical abstract

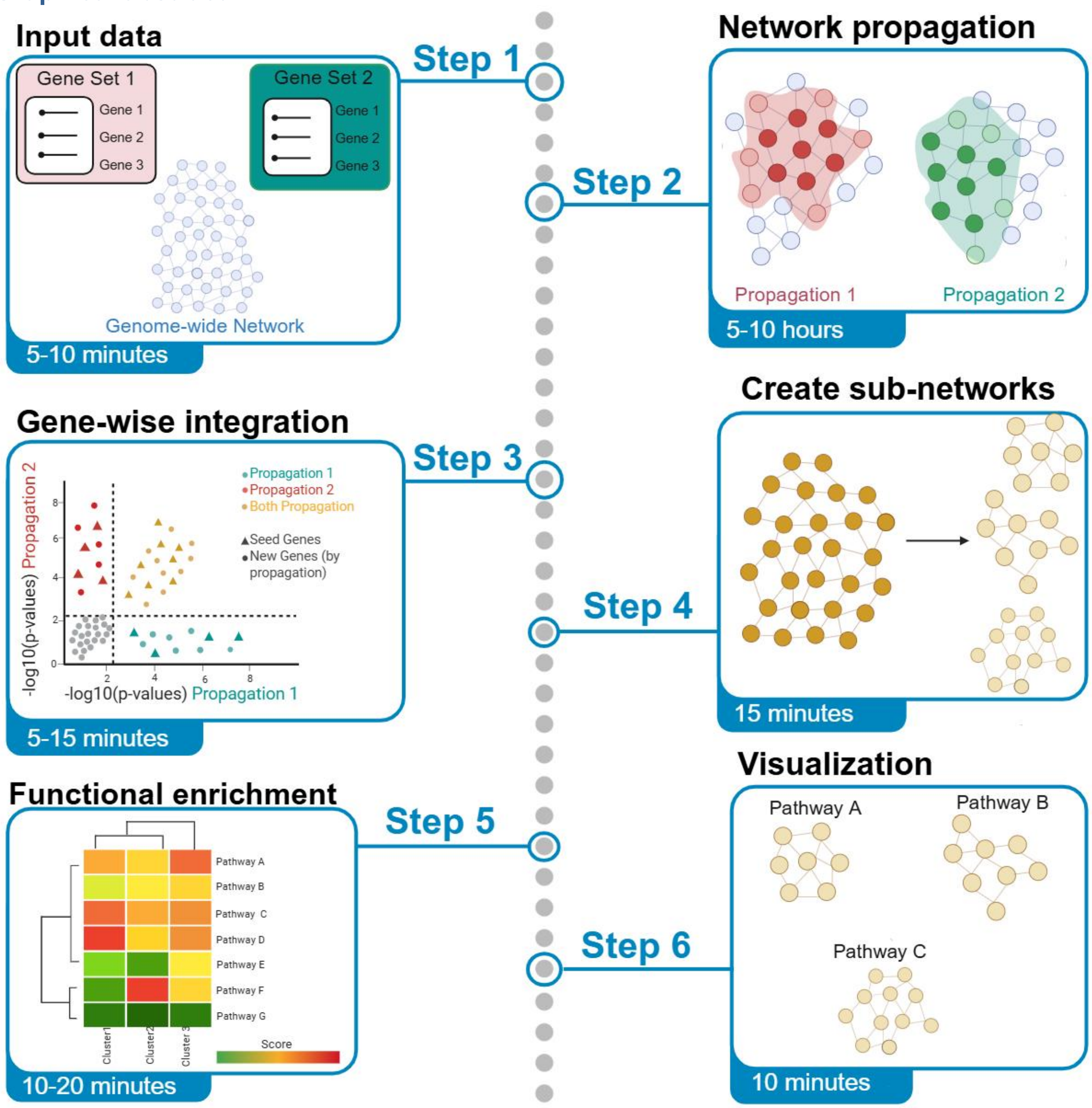

## Before you begin

Integration of data, particularly omics datasets, is critical for understanding mechanistic insights and discerning the interconnectedness of molecular networks underlying health and disease. For instance, an extracellular stimulus can activate a receptor, propagate intracellular signaling through phosphorylation, activate transcription factors, induce gene expression, and ultimately alter protein abundance. Integrating multi-omics measurements is therefore essential for reconstructing the chain of molecular and mechanistic events that links a biological perturbation to a downstream phenotype, particularly given the multilayered nature of cellular signaling pathways. To this end, several approaches have been developed to integrate multi-omics datasets (reviewed in[3]). Owing to the high-dimensionality, heterogeneity, and sparsity of most omics datasets,[4–6] their integration remains a challenge (reviewed in[4,5]). Statistical and computational methods, such as machine learning models, have gained traction in the integration of omics datasets as they are adapted to handling heterogeneous, high dimensional and sparse data (reviewed in[3]). However, some of these methods often require resource-intensive computational infrastructure and elaborate training and often do not integrate decades of prior knowledge on molecular interactions.

A simple computational data integration approach called “network propagation” can integrate multi-omics datasets within the context of prior knowledge molecular interaction networks. Networks are composed of nodes, often representing individual genes in the field of biology, and edges (i.e., connections), representing physical, functional, or other types of associations between genes. We refer to the prior knowledge network as the “base network”. Network propagation is anchored on an assumption that genes implicated in a certain phenotype have a higher tendency to associate with one another.[7] Network propagation is an iterative process in which genes of interest are labeled within a network (hereafter referred to as “seed genes”) and this signal is subsequently propagated (i.e., spread) via the edges to nearby nodes. One can imagine this process as a heat diffusion process, in which heat accumulates in heavily interconnected network neighborhoods. The final propagated scores are converted to p-values by comparing them with scores obtained from randomly labeled seed sets and reflect the network proximity of each gene to the observed seeds, enabling identification of genes that are functionally or physically connected to the seed genes.[8]

In this protocol, we detail the steps involved in performing the two most common network propagation processes: random walk with restart (RWR; also known as insulated heat diffusion or personalized PageRank) and diffusion kernel (DK; also referred to as heat kernel)[8,9]. Interestingly, many widely used search engines, including Google, were built on PageRank-based strategies,[10,11] highlighting the power of network propagation approaches to connect and prioritize information across interconnected systems. Both methods work by spreading signal outward from the seed genes across the network and scoring every gene by how much signal reaches it, so that genes sitting close to many seeds rank highly even if they were not members of the seed genes. The two differ in how the signal fades with distance. RWR models a walker that steps between connected genes but periodically jumps back to the seeds, with a restart probability α that sets how far it strays; the resulting signal spread falls off steadily with distance, with the greatest signal accumulating within the neighbors of the seeds.

Network propagation assigns scores to genes by spreading signal from a set of seed genes through an interaction network. Let $A$ denote the adjacency matrix, $D$ the diagonal matrix of node degrees, and $p_0$ the vector encoding the initial seed signal. Both Random Walk with Restart (RWR) and the Diffusion Kernel (DK) can be expressed as $p = Sp_0$, where the propagation operator $S$ determines how signal spreads and decays across the network. For RWR, $W = AD^{-1}$ is the normalized adjacency matrix with

$$S = \alpha[I - (1 - \alpha)W]^{-1},$$

where the restart probability $\alpha$ controls how strongly propagation remains anchored to the seeds. DK instead models heat diffusing from the seeds for a time $t$, using the graph Laplacian (L=D-A), with

$$S = e^{-tL}$$

Thus, while RWR emphasizes progressively longer network paths according to $(1 - \alpha)$, DK controls the spatial scale of propagation through $t$, with diffusion allowing signal to reach genes beyond the immediate neighborhood. Importantly, in both methods $S$ depends only on the network, not on the input seed genes, allowing it to be computed once and efficiently applied to multiple datasets. Finally, propagation scores can be calibrated against scores from random seed sets of equal size to obtain gene-specific empirical $p$-values, identifying genes that receive more signal than expected given their network position.

After propagation of each dataset individually across the base network, our protocol outlines how to integrate each propagation and identify converging gene sets. We additionally review how to identify pathway layers, or subnetworks, within the larger converging network, using gene set overrepresentation analysis (GSOA). This framework is broadly applicable for identifying shared mechanisms across two or more gene sets, regardless of their experimental source or the technology used to generate them.

## Innovation

The advent of high-throughput techniques has led to multi-modal datasets that are often complex, heterogeneous, and sparse. Integration of these datasets is often needed to gain mechanistic insights. Several approaches, including statistical and computational methods, have been adopted for integration of multi-modal data from single or multiple independent studies. However, many of these approaches are limited by high computational demands or the need for large amounts of data.[3] Here, we review methods for employing network propagation, a simple, robust, and computationally efficient approach for integrating diverse datasets. Network propagation can integrate two or more distinct gene lists regardless of their source, measurement technique, or sample size. By propagating each dataset across the same prior-knowledge network, heterogeneous datasets are projected into a common network space, providing a standardized representation that facilitates direct comparison and integration across molecular data types. The protocol is specific for human genes in HUGO gene nomenclature but can be easily adapted to any organism, provided that the base network and seed gene lists use a common gene nomenclature.

## Institutional permissions

Not applicable to the present protocol.

## Install requisite libraries, prepare input seed sets and organize project directory

**Timing: 5-15 min**

1. Install requisite Bioconductor and CRAN R packages

a. Open RStudio.

b. Run the code block below to install requisite packages.

```
# Install required packages
>install.packages(c("BiocManager", "librarian"))

>librarian::shelf(data.table, dplyr, reshape2, igraph, SparseM, Matrix, ggraph,
eulerr, ggplot2, dynamicTreeCut, parallel, pbapply, plyr, qvalue,
clusterProfiler, ComplexHeatmap, msigdbr, ggrepel, scatterpie, biomaRt, STRINGdb,
patchwork)
```

***Note***: The code above installs the latest available versions of the packages and their dependencies.

***Note***: When running the code block, a message may appear in the console asking whether to update all, some, or none of the packages ([a/s/n]:). Type “a” to update all packages.

***Note***: A dialog box may appear asking whether to restart R before installation. Select “Yes” to ensure that all packages and their dependencies are installed correctly.

***Note***: A prompt may appear asking whether to install packages from source when compilation is required. Select "Yes".

c. Comment out the installation commands after successful installation to prevent the packages from being reinstalled each time the analysis is run.

## 2. Prepare the project directory

a. Create a parent directory named Network_analysis_project or any other names.
b. Save Analysis.Rmd in the parent directory.
c. Save network_propagation_pipeline.R in the parent directory.
d. Create a subdirectory named Seed_sets within the parent directory.
e. Create a subdirectory named Helper_scripts within parent directory.
f. Save the helper scripts used for network propagation and pathway enrichment in the Helper_scripts directory.

***Note***: This protocol uses multiple input files and generates data, tables, and figures. Maintaining these files within a structured project directory facilitates organization and reproducibility. Thus, The network_propagation_pipeline.R script automatically creates propagation process-specific subdirectories for output data, tables, and figures.

## 3. Identify seed sets and assess gene-level overlap across datasets

***Note***: Multi-omics datasets are a key source of seed genes for network propagation. Because omics datasets can contain large numbers of genes or proteins, there is need to prioritize a subset of genes or proteins before network propagation. Examples of seed selection strategies include selecting the most frequently mutated genes from genomic data; selecting the top differentially expressed genes or proteins ranked by log2 fold change from transcriptomic or abundance proteomic data; or selecting proteins containing at least one significantly regulated phosphorylation site from phosphoproteomic data.

a. Prepare each seed set as a file containing one gene per row.
b. Save each seed-set file as .txt, .csv, or .tsv.
c. Place all seed-set files in the Seed_sets subdirectory.

***Critical***: Use the same gene nomenclature for all seed sets.

***Note***: For illustration, this protocol uses two sample gene lists. Seed genes can alternatively be obtained from non-omics data sources.

d. Load the seed sets into RStudio and assess their overlap using the code block below.

```
# Set working directory to the folder containing this R Markdown file
>setwd(dirname(rstudioapi::getSourceEditorContext()$path))

# Seed sets
>seed_files <- list.files("Seed_sets", pattern = "\\.(txt|csv|tsv)$", full.names
= TRUE)

# Inspect overlap before network propagation
# Read seed sets
>seed_sets <- lapply(seed_files, function(x) {data.table::fread(x, header =
FALSE)[[1]]})

# Name seed sets from file names
>data_cols <- tools::file_path_sans_ext(basename(seed_files))

# Identify genes shared across all seed sets
>overlap_genes <- Reduce(intersect, seed_sets)

# Display overlap
>cat("Overlap between seed sets\n")
>cat("Number of seed sets:", length(seed_sets), "\n")
>cat("Number of overlapping genes:", length(overlap_genes), "\n\n")

>if (length(overlap_genes) > 0) {
  cat("Overlapping genes:\n")
  print(overlap_genes)
} else {
  cat("No genes overlap across all seed sets.\n")
}

# Visualize overlap
>seed_cols <- grDevices::hcl.colors(length(seed_sets), "Set 2")
>venn <- eulerr::euler(setNames(seed_sets, data_cols))
>plot(venn, quantities = TRUE, fills = list(fill = seed_cols, alpha = 0.8), labels
= data_cols)
```

***Critical***: Ensure that each seed set contains one gene per row and is saved as a .txt, .csv, or .tsv file in the Seed_sets subdirectory.

***Note***: The code above identifies genes shared across all seed sets and visualizes the overlap using a Euler Venn diagram.

***Note***: Network propagation is most useful when the seed sets have limited overlap and the objective is to identify additional genes that may connect or contextualize the input seed genes. If the seed sets show substantial overlap and the objective is to characterize molecular pathways directly represented by the shared genes, network propagation can be bypassed. In this case, set use_propagation = FALSE in the user-defined preferences section of Analysis.Rmd and proceed directly to network construction, clustering, and functional enrichment using the shared genes. If network propagation is appropriate for the analysis, set use_propagation = TRUE in Analysis.Rmd.

4. Set user-defined preferences for network propagation and downstream analyses

**Note:** As explained in the propagation formula, RWR generally produces a more localized propagation signal, whereas DK allows the signal to diffuse more broadly through the network. Consequently, the two propagation processes can produce different numbers and sets of significant genes. Although users are encouraged to select one propagation process and one base network for their network propagation project, they can also run both propagation processes and compare the resulting analyses.

a. Review the available user-defined parameters in Table 1.

**Table 1. User-defined parameters.**

| Parameter | Options / range | Description |
|---|---|---|
| use_propagation | TRUE/FALSE | Determines whether network propagation is performed. When FALSE, overlapping genes are used directly for network construction, clustering, and functional enrichment. |
| run_pr_check | TRUE/FALSE | Determines whether RWR restart probability (pr) values ranging from 0.1 to 0.9 are evaluated to assist in selecting an appropriate pr value. When FALSE, the user can specify the desired pr value directly. |
| propagation_process | RWR, DK | Specifies the network propagation method. |
| pr | 0.1–0.9, user-defined* | Specifies the RWR restart probability used for network propagation. When run_pr_check = TRUE, pr values ranging from 0.1 to 0.9 are evaluated; when run_pr_check = FALSE, the user can specify the desired pr value directly. |
| integ_operator | sig_all_pval, Min_pval | Specifies how propagated datasets are integrated based on significance levels |
| integ_threshold | 0-0.05 | Specifies how propagated datasets are integrated based on statistical significance |
| network_type | PathwayCommons, STRING | Specifies the interaction network used for propagation. |
| string_score_threshold | 0-1000, user-defined | STRING interaction confidence threshold |
| pc_interaction_types | interacts-with, in-complex-with, or both | Pathway Commons interaction types retained for network analysis |
| use_latest_network | TRUE/FALSE | Determines whether the latest available Pathway Commons or STRING network version is used |
| network_version | User-defined | Specific Pathway Commons or STRING network version when use_latest_network = FALSE |
| num_perms | ≥1, user-defined | Specifies the number of permutations used to assess propagation significance |
| minclustersize | ≥3, user-defined | Specifies the minimum number of genes required for a network cluster |
| enrichment_database | C2, GOBP | Functional enrichment database: canonical pathways (C2) or Gene Ontology Biological Process (GOBP) |
| figures_to_generate | Figure_1, Figure_2, Figure_3 | Specifies which analysis figures are generated |
| save_figures | TRUE/FALSE | Determines whether generated figures are saved to the results directory |
| preview_figures | TRUE/FALSE | Determines whether generated figures are displayed for visual inspection during the analysis |

* Applicable to RWR only; not applicable to DK.

***Note***: For Pathway Commons, the protocol uses "interacts-with" and "in-complex-with" to retain physical interactions by default. Users can modify pc_interaction_types to select other Pathway Commons interaction categories when appropriate.

b. Using information from Table 1, set the user preferences as needed in the code block below.
c. Run the code block to apply the selected preferences to the downstream steps.

```
# Decide analysis strategy
>use_propagation <- TRUE

# Decide whether to evaluate various pr values
>run_pr_check <- FALSE

# Propagation process
>propagation_process <- "RWR"

# RWR restart probability
>pr <- 0.1

# Integration
>integ_operator <- "sig_all_pval"
>integ_threshold <- 0.05

# Network
>network_type <- "PathwayCommons"

>use_latest_network <- TRUE
# network_version <- "v14"

>string_score_threshold <- 900

>pc_interaction_types <- c("interacts-with", "in-complex-with")

# Clustering and enrichment
>num_perms <- 10000
>minclustersize <- 10
>enrichment_database <- "GOBP"

# Figure controls
>figures_to_generate <- c("Figure_1", "Figure_2", "Figure_3")
>save_figures <- TRUE
>preview_figures <- TRUE
```

***Note***: The settings shown above correspond to the default settings used to generate the sample outputs for this protocol.

***Note***: The sample outputs were generated using physical interactions from Pathway Commons network version 14, which was the latest available version when the analysis was performed.

5. Evaluate the composition of significant propagated genes across different RWR restart probabilities

**Timing: approximately 20–72 hours, depending on network, seed sets, and number of permutations.**

***Note***: Evaluating multiple pr values requires a separate network propagation for each seed set and each pr value and can therefore be computationally time-consuming. For routine analyses, pr values between 0.1 and 0.3 can be used without performing the optional pr check. The pr checking procedure evaluates values from 0.1 to 0.9 and generates a plot showing the composition of significant propagated genes as seed genes versus newly identified genes across the evaluated values.

a. Ensure run_pr_check and use_propagation are set toTRUE in the user-defined preferences code block above.
b. Run the pr checking procedure shown in the code block below

```
>if (run_pr_check) {
  source("Helper_scripts/PR_checking.R")
  pr_check <- run_pr_checking(
    seed_files = seed_files,
    network_type = network_type,
    pr_values = seq(0.1, 0.9, by = 0.1),
    num_perms = num_perms,
    use_latest_network = use_latest_network,
    network_version = if (exists("network_version")) network_version else NULL,
    string_score_threshold = string_score_threshold,
    pc_interaction_types = pc_interaction_types
  )
    print(pr_check$plot)
}
```

***Note***: The PR_checking.R helper script is automatically sourced from the Helper_scripts directory when run_pr_check = TRUE in the user-defined preferences code block. The resulting plot shows the relative composition of significant propagated genes as seed genes and newly identified genes across the evaluated pr values.

## Key resources table

| REAGENT or RESOURCE | SOURCE | IDENTIFIER |
|---|---|---|
| **Software and algorithms** | | |
| R v4.6.1 | R Core Team | https://www.r-project.org/ |
| RStudio v2026.07.1+147 | RStudio Team | https://docs.posit.co/ide/user/ |
| librarian v1.8.1 | CRAN | https://cran.r-project.org/package=librarian |
| data.table v1.18.4 | CRAN | https://cran.r-project.org/package=data.table |
| reshape2 v1.4.5 | CRAN | https://cran.r-project.org/package=reshape2 |
| dplyr v1.2.1 | CRAN | https://cran.r-project.org/package=dplyr |
| igraph v2.3.3 | CRAN | https://cran.r-project.org/package=igraph |
| SparseM v1.84-2 | CRAN | https://cran.r-project.org/package=SparseM |
| Matrix v1.7-5 | CRAN | https://cran.r-project.org/package=Matrix |
| ggraph v2.2.2 | CRAN | https://cran.r-project.org/package=ggraph |
| eulerr v8.1.0 | CRAN | https://cran.r-project.org/package=eulerr |
| dynamicTreeCut v1.63-1 | CRAN | https://cran.r-project.org/package=dynamicTreeCut |
| ggplot2 v4.0.3 | CRAN | https://cran.r-project.org/package=ggplot2 |
| pbapply v1.7-4 | CRAN | https://cran.r-project.org/package=pbapply |
| plyr v1.8.9 | CRAN | https://cran.r-project.org/package=plyr |
| ggrepel v0.9.8 | CRAN | https://cran.r-project.org/package=ggrepel |
| scatterpie v0.2.6 | CRAN | https://cran.r-project.org/package=scatterpie |
| Msigdbr v26.1.1 | CRAN | https://cran.r-project.org/package=msigdbr |
| ComplexHeatmap v 2.28.0 | Bioconductor | https://bioconductor.org/packages/ComplexHeatmap |
| qvalue v2.44.0 | Bioconductor | https://bioconductor.org/packages/qvalue |
| clusterProfiler v4.20.0 | Bioconductor | https://bioconductor.org/packages/clusterProfiler |
| STRINGdb v2.24.0 | Bioconductor | https://bioconductor.org/packages/STRINGdb |
| biomaRt v2.68.0 | Bioconductor | https://bioconductor.org/packages/biomaRt |
| Other | | |
| Intel(R) Core(TM) i7-10610U CPU @ 1.80GHz (2.30 GHz) with 32 GB RAM | NA | NA |

## Materials and equipment

Install the latest R version and RStudio on a >=16 GB RAM computer. Install all the packages in the Key Resources Table as well as their dependencies. The entire protocol can be executed in most Windows and macOS computers (hardware used is listed in Key Resources Table).

## Step-by-step method details

**Timing: 5-35 hours, variable depending on the propagation process, base network used**

The workflow is executed through the user-facing Analysis.Rmd file and supporting helper scripts. User-defined parameters are specified in Analysis.Rmd, while the helper scripts are called automatically.

### 6. Run the network analysis. Troubleshoot 1-3

a. Ensure that an internet connection is available before the first analysis run.

***Note***: An internet connection is required to download the selected interaction network the first time it is used.

b. Source the main helper script.
c. Run the network analysis.

```
>source("network_propagation_pipeline.R")

>results <- run_network_propagation(
  seed_files = seed_files,
  use_propagation = use_propagation,
  propagation_process = propagation_process,
  network_type = network_type,
  integ_operator = integ_operator,
  integ_threshold = integ_threshold,
  num_perms = num_perms,
  minclustersize = minclustersize,
  enrichment_database = enrichment_database,
  pr = pr,
  use_latest_network = use_latest_network,
  network_version = if (exists("network_version")) network_version else NULL,
  string_score_threshold = string_score_threshold,
  pc_interaction_types = pc_interaction_types
)
```

***Critical***: Ensure that the seed gene lists and the base network use the same gene nomenclature to ensure that the seed genes are correctly matched to the network.

***Note***: The main helper script automatically performs network acquisition and preprocessing, network propagation, integration of propagated signals, network construction and clustering, functional enrichment, and generation of analysis outputs, including figures and tables.

***Note***: **Network acquisition and preprocessing**. The main helper script acquires and preprocesses the selected base interaction network. The workflow currently supports Pathway Commons and STRING, with the option to download the latest available network or a specified version according to the user-defined settings. Table 2 summarizes the characteristics of commonly used biological interaction networks.

***Note***: **Network propagation**. When use_propagation = TRUE, the pipeline performs network propagation separately for each seed set. Network propagation consists of calculation of the inverse/Laplacian matrix, propagation of the molecular signal through the network, permutation testing, and calculation of gene-level p values. Separate helper scripts implement these operations for the selected propagation process. The resulting gene-level p values quantify the statistical proximity of genes to the seed genes.

***Note***: **Integration of propagated signals**. The propagation results generated for the individual seed sets are integrated gene-wise based on their p values. With sig_all_pval, genes must be significant across all separate seed-set propagations; with Min_pval, genes are retained if they are significant in at least one propagation. The significance threshold can range from 0 to 0.05. sig_all_pval provides a more stringent selection, whereas Min_pval produces a more inclusive gene set (Table 3). The selected integrated gene list is used for downstream network construction. For illustration, the protocol uses sig_all_pval.

***Note***: **Network clustering**. Clustering a network into subnetworks entails partitioning the network into smaller, functionally related subnetworks (i.e., clusters). The main helper script implements a custom R-based network clustering method.[1] The approach performs two layers of network clustering. In the first layer, the network is clustered using the cluster_walktrap function from the igraph R package. This method simulates short random walks that tend to remain within densely connected regions of the network, thereby grouping genes that are repeatedly visited together. In the second layer, the resulting clusters are further refined using the cutreeDynamic function from the dynamicTreeCut R package. Specify the minimum cluster size using minclustersize according to the desired resolution of the analysis. A smaller minimum cluster size tends to produce more cluster pieces, whereas a larger value tends to produce fewer clusters. As a rule of thumb, evaluate several minimum cluster size values and select a value at which the network clusters are stable, i.e., increasing the minimum cluster size further no longer causes clusters to disintegrate into smaller pieces.

***Note***: **Functional enrichment**. The clustered networks are analyzed using GSOA with MSigDB gene-set collections, including Canonical Pathways (C2) and Gene Ontology Biological Process (C5). The clustering procedure is intended to identify smaller subnetworks associated with distinct biological processes; consequently, individual clusters are expected to show enrichment for distinct functional terms.

**Note**: When use_propagation = FALSE, the workflow bypasses the propagation and integration steps and directly constructs and clusters a network from genes shared across the seed sets. Functional enrichment is then performed on the resulting clustered networks.

**Table 2. Common biological networks and their key features.**

| **Network** | **Description** | **Interaction type** | **Edge weights** | **Interaction-type annotation** | **Organisms covered** | **Source** |
|---|---|---|---|---|---|---|
| Pathway Commons | Integrated pathway and molecular interaction knowledge from BioPAX. | Physical & functional | No | Yes | Human | https://www.pathwayc |
| ConsensusPathDB | Integrated protein interactions, regulatory interactions, and biochemical pathways from multiple databases. | Physical & functional | Partial | Yes | Human, mouse, yeast | http://cpdb.molgen.mp |
| FunCoup | Bayesian functional coupling network integrating diverse genomic evidence. | Physical & functional | Yes | Yes | 22 model organisms | https://funcoup.org/ |
| GeneMANIA | Functional association network integrating co-expression, interactions, pathways, and other evidence. | Physical & functional | Yes | Yes | 9 organisms | https://genemania.org/ |
| HumanBase | Tissue- and cell-type-specific human functional gene networks. | Functional | Yes | No | Human | https://hb.flatironinstitu |
| HumanNet | Integrated human functional gene network based on genomic and proteomic evidence. | Physical & functional | Yes | No | Human | https://www.inetbio.org |
| STRING | Protein association network integrating experimental, curated, and predicted evidence. | Physical & functional and others | Yes | Yes | >12,000 organisms | https://string-db.org/ |
| CORUM | Curated mammalian protein complexes. | Physical | No | No | Primarily human, mouse, and rat | https://mips.helmholtz-muenchen.de/corum/ |
| Reactome | Functional interaction network derived from Reactome pathways and interaction data. | Physical & functional | No | Yes | Human; ~20 other species via computational orthology inference | https://reactome.org/ |
| BioGRID | Curated physical and genetic interactions from experimental literature. | Physical & functional | No | Yes | Human and >70 model organisms | https://thebiogrid.org/ |

**Table 3. Integration approaches for combining propagated signals across seed sets**

| | **sig_all_pval** | **Min_pval** |
|---|---|---|
| Selection rule | Gene is significant in **all** seed-set propagations | Gene is significant in **at least one** seed-set propagation |
| Stringency | Higher | Lower |
| Threshold range | 0–0.05 | 0–0.05 |
| Typical result | Smaller, high-confidence gene set | Larger, more inclusive gene set |

### 7. Inspect and adjust analysis outputs. Troubleshoot 4.

Run the code block below to preview the selected figures and, if needed, adjust their dimensions and save the revised figures.

```
>generate_figures(
  results = results,
  save_figures = save_figures,
  preview_figures = preview_figures,
  figures = figures_to_generate,

  figure_1A_width = 3.7,
  figure_1A_height = 3.0,
  figure_1B_width = 3.7,
  figure_1B_height = 1.5,
  figure_1C_width = 3.2,
  figure_1C_height = 1.5,

  figure_2_width = 11.69,
  figure_2_height = 8.27,
  figure_3_width = 5.6,
  figure_3_height = 3.57
)
```

## Expected outcomes

Network propagation expands a set of seed genes through a molecular interaction network to identify genes in their network neighborhood and molecular pathways that may not be apparent from the original gene set alone.

Integration of datasets using network propagation is particularly useful when two or more gene sets have limited overlap at the gene level (Figure 1A) but may converge at the molecular pathway level. Before launching network propagation, users should assess (i) the extent of gene-level overlap between the seed sets and (ii) whether the overlapping genes form a network suitable for clustering and functional enrichment. When gene-level overlap is limited, network propagation can identify additional genes that connect the seed sets within the interaction network.

Network propagation can also be used when mechanistic insights cannot be obtained from a single gene set. In such cases, a gene set of interest can be integrated with relevant gene sets derived from external resources, such as cBioPortal for Cancer Genomics (https://www.cbioportal.org/), DisGeNET (https://disgenet.com/), the Catalogue of Somatic Mutations in Cancer (COSMIC; https://www.sanger.ac.uk/tool/cosmic/), or genome-wide association studies, to identify molecular pathways potentially shared between the datasets.

As demonstrated in this protocol, integration of distinct gene sets can generate propagated and integrated networks, clustered subnetworks, and functional enrichment results highlighting molecular pathways or biological processes shared by the datasets of interest (Figures 1–3, Table S1–S3). The propagation results also show the relative yield of significant seed genes and newly identified genes, providing an indication of the extent of network expansion (Figure 1B–C). The full and clustered networks are generated side by side (Figure 2), and the functional enrichment map/heatmap is generated for the clustered networks (Figure 3). The network edge list and node table are additionally saved in the Tables directory and can be exported for visualization in Cytoscape. Common observations, potential causes, and recommended actions for troubleshooting network analysis outcomes are summarized in Table 4. These results can generate testable hypotheses and identify molecular pathways for validation in follow-up functional assays.

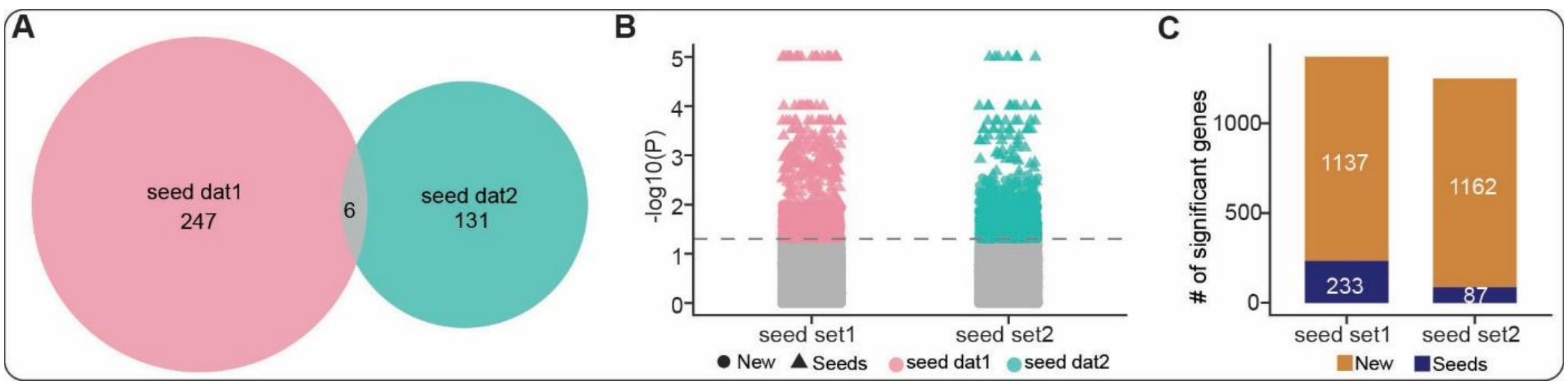

**Figure 1. Examples of expected outputs from the generate_figures function.**
(**A**) Venn diagram depicting the overlap between genes in seed set1 and seed set2.
(**B**) Jitter plot showing the categories of genes identified by propagating seed set 1 and 2, including seed genes and newly identified genes, stratified by statistical significance ($p < 0.05$).
(**C**) Bar plots showing the number of significant seed genes and newly identified from the individual propagation analyses. Network propagation output dataset used to generate Figure 1B-C is provided in Table S1.

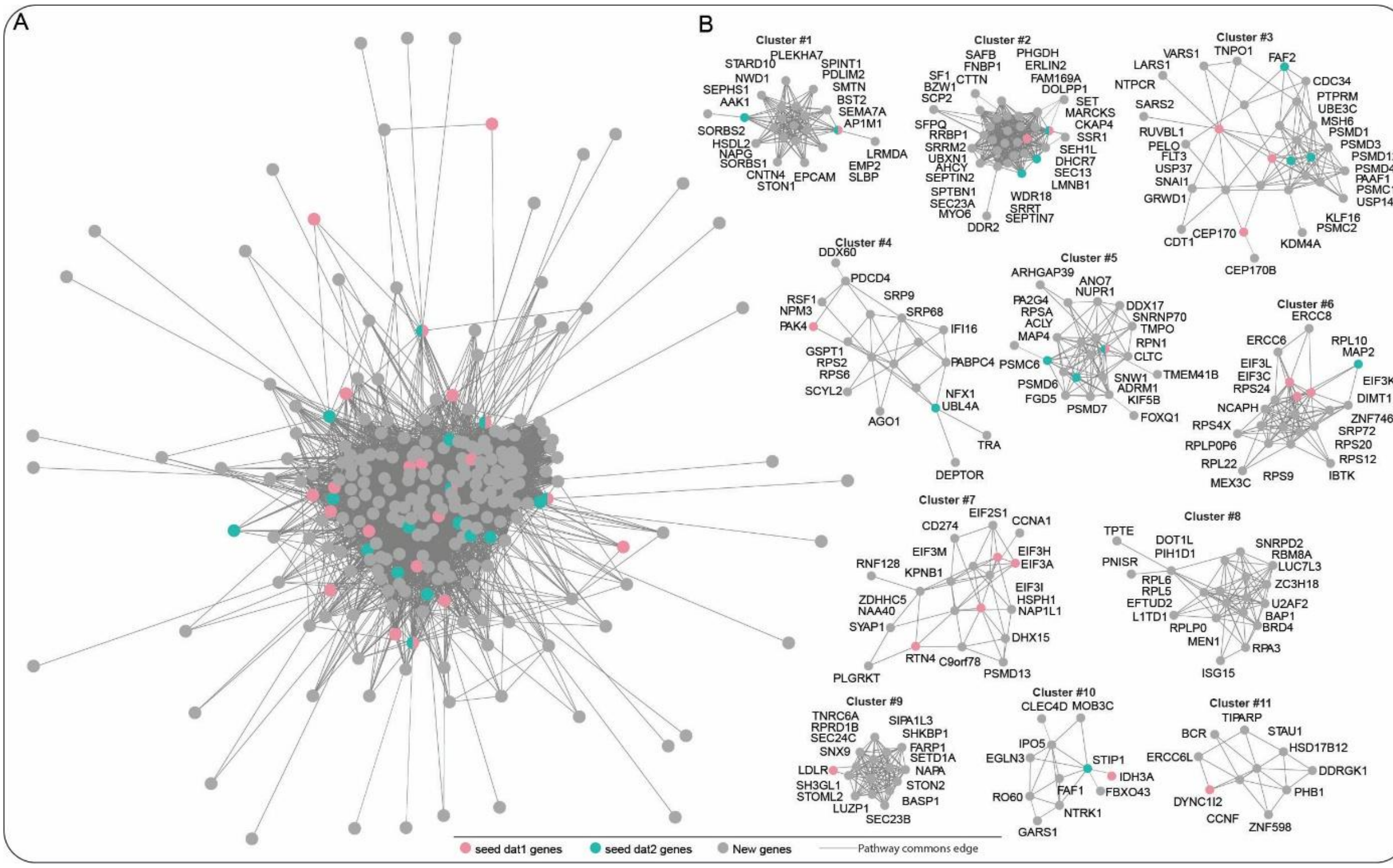

**Figure 2. Examples of converging networks and subnetwork clusters.**
(**A**) Full network generated from genes significant in both seed set 1 and seed set 2 propagation analyses. Edge list for the full network is provided in Table S2A.
(**B**) Clustered network generated from the full network using a custom R-based clustering approach. See edge list for the clustered network (Table S2B) and node table (Table S2C) provided.

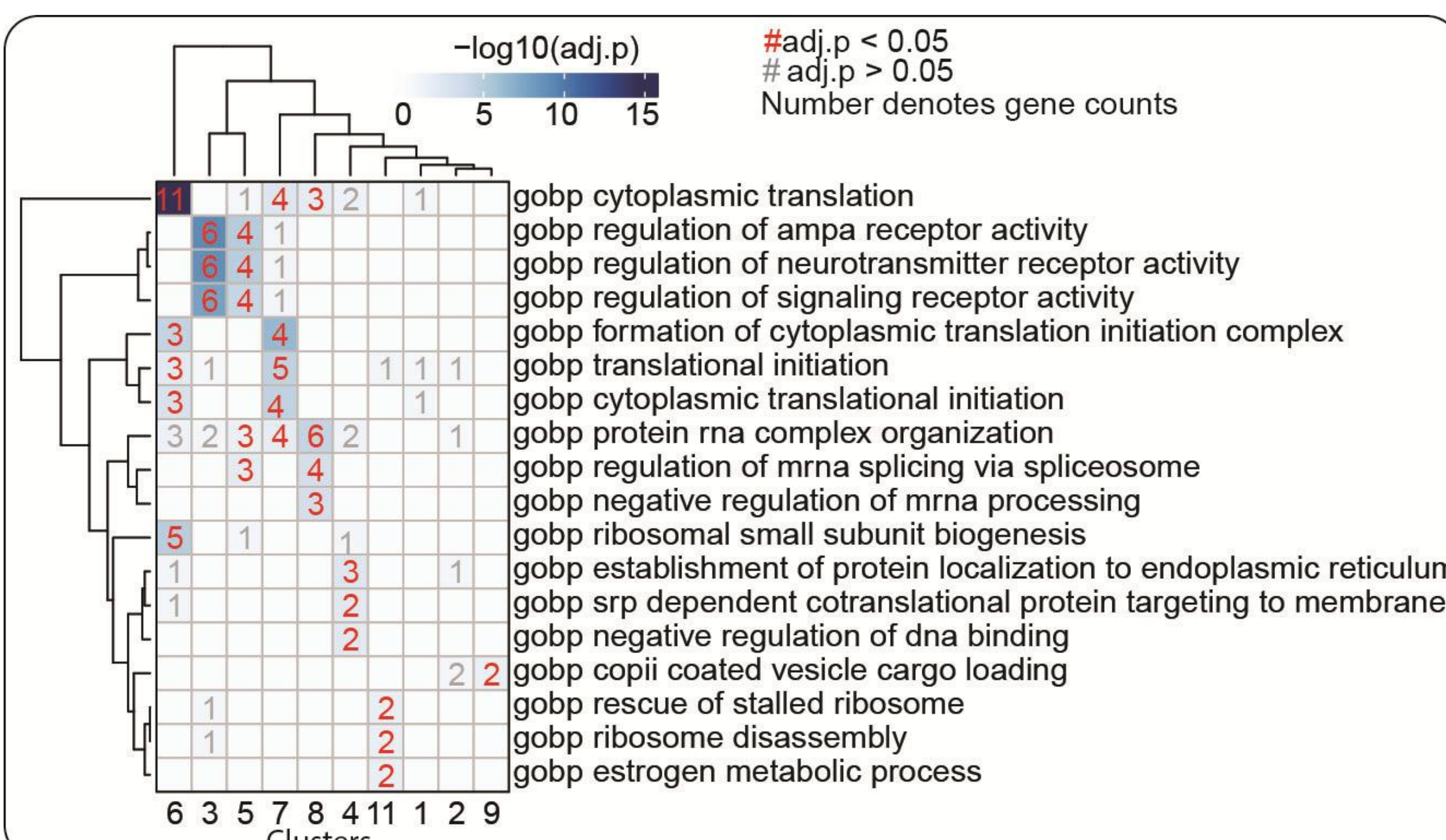

**Figure 3. Example of GSOA for subnetwork clusters.** The p-values were calculated by a hypergeometric test with multiple hypothesis testing correction (i.e., false discovery rate). The top 3 significantly enriched biological process terms are visualized per cluster. Numbers denote the number of genes intersecting between the pathway and the cluster, with red indicating significant enrichment (adjusted p-value < 0.05). GSOA output is for clusters provided in Figure 2. A complete enrichment table is provided as Table S3.

**Table 4. Interpretation of network analysis outcomes and recommended actions**

| **Observation / problem** | **Possible cause** | **Recommended action** |
|---|---|---|
| Few or no genes pass the propagation significance threshold | Seed set may be too small, network connectivity may be low, or significance threshold may be stringent | Confirm that seed genes are correctly formatted and present in the selected network; consider increasing num_perms or reviewing integ_threshold. |
| Most significant genes are from the original seed sets | Limited propagation beyond the input genes | Review seed-set size and network connectivity; consider a different network or propagation parameters. |
| Little or no overlap between seed sets | Seed sets may represent distinct biological signals | Consider whether network propagation is appropriate; inspect the overlap before proceeding. |
| No overlapping genes are found in the selected network | Input genes may be absent from the network or use incompatible gene identifiers | Check gene identifiers and confirm that the selected network supports the organism under study. |
| Direct analysis (use_propagation = FALSE) produces a small network | Few genes are shared across the seed sets or shared genes have limited network connectivity | Confirm the seed-set overlap and consider whether a different interaction network is appropriate. |
| Network contains several small/disconnected components | Input genes or propagated genes have limited connectivity in the selected network | Review the selected network and minclustersize; consider whether the resulting network is biologically interpretable. |
| No network clusters are identified | Network is too small or lacks sufficient connectivity for clustering | Review the number and connectivity of network genes and consider reducing minclustersize where appropriate. |
| Enrichment analysis returns few or no significant terms | Network clusters may be small or lack sufficient functional coherence | Review cluster sizes and consider the alternative enrichment database (C2 or GOBP). |
| Figure is crowded or difficult to interpret | Network size or number of clusters is large | Adjust the figure dimensions in generate_figures() and/or inspect individual clusters using the exported node and edge tables. |
| Figures are not displayed or saved | preview_figures or save_figures is set to FALSE, or the output directory is unavailable | Confirm the figure-control parameters and verify that the results directory is writable. |

## Limitations

Although the network propagation processes described in this protocol are robust and can give mechanistic insights, these processes possess some limitations. First, due to the extent to which the signal from the seed genes is spread across a base network, the molecular pathways or biological processes resulting from the two network propagation processes may vary. Due to the level of connections between nodes in different base networks, some variations may be seen between different base networks. Identifying terms that converge across network propagation processes and/or base networks could strengthen the findings. Secondly, network propagation analysis produces several significant molecular pathways or biological process terms making it challenging to select the most relevant term. Thus, the choice of the molecular pathway(s) subjected to follow up wet lab validation may be subjective. It may be useful to rank the terms and identify at least the top three for downstream validation. Ranking subnetworks using some quantitative information (e.g., mutation rates, protein abundance, gene expression etc.) from experimental studies or databases could minimize subjectivity, as done in our prior work.[1] Third, this approach is limited to measurement technologies whose outputs can be mapped onto the entities represented in the underlying network, most commonly genes or proteins. Thus, datasets that cannot be readily linked to these entities (e.g., lipids or metabolites, unless they are somehow connected to regulatory enzymes) may be more difficult to integrate using this framework. Lastly, the choice of a base network for a set of seed genes may be challenging. It is advisable to understand the properties of the base networks and the goal of the study to guide in selection of the base network. For example, a study that is more interested in physical or functional interactions should choose a base network that is well known to have the type of interactions of interest (Table 2).

## Troubleshooting

### Problem 1: Users encounter a sparse-to-dense coercion warning or error when calculating the inverse or Laplacian matrix. Related to Troubleshoot 1.

Users may see the following warning or error during network propagation:
Warning: asMethod(object) : sparse->dense coercion: allocating vector of size 3.0 GiB when calculating the inverse or Laplacian matrix.
Error: asMethod(object) : sparse->dense coercion: allocating vector of size 3.0 GiB when calculating the inverse or Laplacian matrix.

**Potential solution:**
The warning indicates that R is attempting to convert a sparse matrix to a dense matrix and may require substantial memory. If the message appears as a warning and the analysis continues, no action is required. If the operation terminates with an error, increase the memory limit available to R and restart R before rerunning the analysis. For example:
Sys.setenv(R_MAX_VSIZE = "100Gb")
The value should be adjusted according to the available system memory.

### Problem 2: Users encounter an HTTP request or Ensembl connection error while mapping network identifiers. Related to Troubleshoot 2.

Users may encounter errors such as:
Error in `req_perform()`:
! Failed to perform HTTP request.
Caused by error in `curl::curl_fetch_memory()`:
! Could not resolve hostname [www.ensembl.org]
or:
Error in httr2::req_perform():
! HTTP 405 Method Not Allowed.

**Potential solution:**
These errors generally occur when the connection to the Ensembl server is unavailable or unstable. Ensure that the computer has a stable internet connection and rerun the affected step. If the error persists, retry the analysis later.

Problem 3: Users encounter timeout or incomplete-download errors when retrieving Ensembl or STRING network files. Related to Troubleshoot 3.
Users may encounter errors such as:
Error in `httr2::req_perform()`:
! Failed to perform HTTP request.
Caused by error in `curl::curl_fetch_memory()`:
! Timeout was reached
or:
Warning in download.file(...):
downloaded length ... != reported length ...
Error in download.file(...):
download ... failed

**Potential solution:**
These errors may occur when downloading large network files over a slow or unstable internet connection. Increase the R download timeout before rerunning the affected Analysis.R:
options(timeout = 600)
The timeout can be increased further if necessary.

Problem 4: Figure generation fails or figures are not displayed/saved. Related to Troubleshoot 4.
**Potential solution:**
Check that the required plotting packages are installed, that save_figures and preview_figures are set as intended, and that the results directory is writable. If a figure is too large or crowded, increase the corresponding width and height parameters in generate_figures() and rerun the figure-generation step.

Problem 5. User encounter an error related the Figures. Related to Troubleshoot 4.
Error in grDevices::pdf(file.path(fig_dir, filename), width = width, height = height, : cannot open file 'results/Figures/RWR/RWR_PathwayCommons_Figure_xx.pdf' Called from: grDevices::pdf(file.path(fig_dir, filename), width = width, height = height, useDingbats = FALSE)

**Potential solution:**
This error may occur when one or more PDF figure files are open while Analysis.R is being run. Close any open PDF figure files and rerun the affected Analysis.R chunks.

## Resource availability

### 1. Lead contact

Further information and requests for resources and reagents should be directed to and will be fulfilled by the lead contact, Mehdi Bouhaddou (bouhaddou@ucla.edu).

### 2. Technical contact

Technical questions on executing this protocol should be directed to and will be answered by the technical contact, Charles O. Olwal (charlesolwal26@ucla.edu).

## 3. Materials availability

This study did not generate new materials.

## 4. Data and code availability

The datasets generated and/or analysed during the current study and the codes are available in the GitHub repository, https://github.com/BouhaddouLab/MONeP-Multi-Omics-Network-Propagation and its supplementary files.

## ACKNOWLEDGMENTS

We acknowledge the UCLA AIDS Institute, the James B. Pendleton Charitable Trust, and the McCarthy Family Foundation for their generous support of our research. We additionally would like to acknowledge funding from the National Institutes of Health (NIH-NIGMS R35GM160071 to MB). Our trainees were generously supported by the Howard Hughes Medical Institute (HHMI) Gilliam Fellowship (#GT17454 to SKM), Molecular Biology Institute Whitcome Fellowship to SKM, the UCLA-Caltech Medical Scientist Training Program (NIH-NIGMS T32GM008042 to SFB), the David Geffen Medical Student Scholarship to SFB, the Cellular and Molecular Biology T32 at UCLA (NIH-NIGMS T32GM145388 to DMW), the Microbial Pathogenesis Training Grant (MPTG NIH-NIAID T32AI007323 to YD), and the Graduate Research Fellowship Program (GRFP) from the National Science Foundation (DGE-2444110 to YD).

## AUTHOR CONTRIBUTIONS

Conceptualization of the study: COO, MB. Writing-original draft: C.O.O. Investigation: COO, DMW, IA A, VV, AR, ALD, KFD, YD, SKM, SFB, SS. Supervision: MB. Funding acquisition: MB. Writing-review: DMW, IA A, VV, AR, ALD, KFD, YD, SKM, SFB, SS, MB.

## DECLARATION OF INTERESTS

The authors declare no competing interests.

## Declaration of generative AI and AI-assisted technologies in the writing process

Portions of the texts and codes were refined with assistance from ChatGPT. All content generated was then reviewed, edited and approved by the authors, who take full responsibility for the final paper.

## Table captions

Table S1: Network propagation outputs for various seed set, related to Figure 1B-C.
Table S2. Network edge list and nodes. (A) Edge list for the full network, (B) edge list for the clustered network and (C) node table for the clustered network, related to Figure 2.
Table S3: Enrichment output for the gene set overrepresentation enrichment analysis of the node table above as the input, related to Figure 3.